**Polymer-based composites for engineering organic memristive devices**

*Carlos David Prado-Socorro, Silvia Giménez-Santamarina, Lorenzo Mardegan, Luis Escalera-Moreno, Henk J. Bolink, Salvador Cardona-Serra\* and Eugenio Coronado*

Institute of Molecular Science, Universitat de València, C/ Catedrático Jose Beltrán nº2, 46980 Paterna, Valencia, Spain

E-mail: salvador.cardona@uv.es

Keywords: organic electronics, electrochemical cells, memristive materials, organic semiconductor, molecule-based materials, electroactive materials, memristors, polymeric materials.

Memristive materials are related to neuromorphic applications as they can combine information processing with memory storage in a single computational element, just as biological neurons. Many of these bioinspired materials emulate the characteristics of memory and learning processes that happen in the brain. In this work, we report the memristive properties of a two-terminal (2-T) organic device based on ionic migration mediated by an ion-transport polymer. The material possesses unique memristive properties: it is reversibly switchable, shows tens of conductive states, presents Hebbian learning demonstrated by spiking time dependent plasticity (STDP), and behaves with both short- (STM) and long-term memory (LTM) in a single device. The origin and synergy of both learning phenomena were theoretically explained by means of the chemical interaction between ionic electrolytes and the ion-conductive mediator. Further discussion on the transport mechanism was included to explain the dynamic behaviour of these ionic devices under a variable electric field. We propose this polymer-based composite as an outstanding neuromorphic material for being tunable, cheap, flexible, easy to process, reproducible, and more biocompatible than their inorganic analogues.

1. Introduction

Current modern computers are built under the traditional von Neumann architecture, which is characterized by a memory-computing structure where the memory unit and the central processing unit (CPU) are physically separated. Under this architecture, the memory unit stores both data and program instructions, sharing the same data-transfer channel to the CPU.[1] The consequence is that the processor is idle for a certain amount of time while memory is accessed. This limitation is known as von Neumann bottleneck.[2] Such non-desirable effect critically limits the effective processing speed of large amounts of data, hindering the development of future AI applications such as pattern recognition and big data mining, among others. To mitigate this limitation, a variety of proposals based on non-von Neumann architectures are being explored by the scientific community. Mostly based on the idea of unifying processing and memory units, in-memory processing technologies have emerged as the most promising solution. These technologies attempt to mimic the most efficient computer in terms of processing speed, learning processes and low energy consumption: the biological brain (see **Figure 1a**). With around 20W of power consumption, it intensively processes and classifies daily life big-data, a feat that no artificial proposal has managed to achieve up to date.[3]

In this regard, one of the most promising ideas consists in using a novel electronic component: the memristor. Theoretically proposed by Chua *et al*. in 1971,[4,5] a memristor is a fundamental electronic element at the level of resistors, inductors, and capacitors, that portrays the ability of varying its intrinsic resistance depending on the previous history of voltage applied to it. Materials showing this capability are still under exhaustive exploration for developing functional memristive devices. Initially, such devices were suggested as resistive memories (ReRAM) for digital storage of information. However, their full potential is achieved in analogue computing,[6] where the continuous variation of resistance shown by memristive devices can accurately be used to solve complex 'maze-type' problems.[7-9]

The first physical characterization of a memristive device was reached with the $TiO_2$ memristor.[10] The mechanism is based on oxygen vacancies electroformed at the surface of a titanium dioxide layer sandwiched by two titanium electrodes.[11,12] Soon after, a variety of alternatives for obtaining new memristive mechanisms emerged. Fruitful remarkable examples include phase change memories (PCM),[13-15] metallic filament migration,[16-18] and magnetic tunnel junctions (MTJ);[19-22] all of them using solid-state inorganic materials like noble metals or metal oxides. Despite the recent success of these extended inorganic materials, no large-scale commercialization has yet been achieved, mostly because their memristive capabilities are inherently limited by their low chemical variability and functionalization, difficulties in nanostructuration, reduced biocompatibility, and, more importantly, poor reproducibility due to large parameter dependences.[23-25]

However, organic materials can potentially form memristive devices which excel in these aspects, as they take advantage of high versatility in molecular chemistry to design materials while reducing the size and energetic requirements.[26] These materials are also 'green' and may improve the processability by means of low-cost soft techniques. Other advantages include flexibility, low power consumption, dense circuit integration, and biocompatibility, which is crucial for applications in the field of nervetronics.[27] All of these characteristics have set the focus of material chemists towards obtaining new organic memristive devices.[28-35]

One of the first attempts at this was presented by Malliaras *et al*. in 2010.[28] In this contribution, the authors published a bistable electrochemical cell with the $[Ru(bpy)_3(PF_6)_2]$ complex as active material. This device aimed to mimic the structure of the original $TiO_2$ memristor, but benefitting from the ionic redistribution of $PF_6^-$ anions upon application of a bias voltage. Later, in 2014, Lei *et al*. tested a device consisting of a thin layer of polyvinyl alcohol in the absence of ions.[36] However, despite the novelty of including a polymeric material, Lei's work did not introduce any insight into the mechanism governing the observed memristive behaviour.[36] Thus, all attempts to design and improve these devices have remained so far unsuccessful, only showing 'write-once, read-many' devices at best.[37] In addition, these organic devices required the use of a 3-T structure to obtain a relatively long memory remanence.[38-41]

In this work, we overcome such complexity by taking a step forward in the fabrication, design, and learning properties of a polymer-based composite 2-T memristive device. This proof of concept is based on a mixture of an organic conductive polymer, a solid electrolyte and an embedded salt. Including a solid electrolyte is crucial since it benefits our approach in two ways: first, its structure embodies a large concentration of polar groups on the side chains, permitting adequate solvation of the salt, which in turn maximizes the number of ionic carriers. Secondly, the cohesive energy of the polymer is low enough to maintain a certain flexibility, allowing for the reorientation of the local coordination geometry as well as the segmental motion of the side chains, while preferably coordinating the cations. The distinct mobilities of anions and cations are proposed to be the origin of different response to voltage, producing two memory regimes. We characterize the memristive behaviour by means of analogue conductive performance, revealing a continuous scheme of achievable and reversible states. In this way, multiple controllable hysteresis loops in the I-V curves are reported, along with Excitatory Postsynaptic Current (EPSC), Short-Term Memory (STM) and Long-Term Memory (LTM), and Spike-Time Controlled Plasticity (STDP).

2. Results and Discussion

Synaptic devices with a vertical structure (**Figure 1b**) were fabricated by implementing an active layer composed of a mixture of three materials: *poly[{2,5-di(30,70-dimethyloctyloxy)1,4-phenylenevinylene}-co-{3-(40-(300,700-dimethyloctyloxy)phenyl)1,4phenylenevinylene}-co-{3-(30-(30,70-dimethyloctyloxy)phenyl)-1,4-phenylenevinylene}]* (Super Yellow, organic semiconductor), Hybrane® DEO750 8500 (ion-transport polymer), and lithium triflate (LiCF$_3$SO$_3$) (**Figure 1c**). These precursors were dissolved in cyclohexanone at a concentration of 15, 10, and 5 mg/mL, respectively, and then mixed following the mass ratio of 1:0.30:0.09 with respect to Super Yellow, ending with a final concentration of 8.72, 2.62, and 0.78 mg/mL. Pre-patterned indium tin oxide (ITO) coated glass plates were used as transparent conductive substrates. After the cleaning process, the substrate was dried and the prepared solution was spun at 2000 rpm for 60s. The as-coated films were annealed for 3 hours at 75 °C to form a uniform, homogeneous thin layer. The obtained film presented a thickness of 209 nm, characterized by profilometry measurements. Topography images were taken through an AFM for a different spin-coating rotational speed rate (see **Supporting Information S1 and S2**). The top electrode (Ag, 100 nm thickness) was thermally evaporated on top of the active layer using a shadow mask. The active area of the junction is of 0.0825 cm$^2$. The whole fabrication procedure (i.e., preparation of solutions, spin-coating, annealing, and thermal evaporation) has been conducted under N$_2$-filled atmosphere to assure maximum reproducibility while avoiding contamination. The robustness of these kind of active layers has also been proved by incorporating them into light emitting electrochemical cells, which have shown long electroluminescence lifetime (~1600h).[42]

This chemically engineered structure was then electrically characterized in order to unveil the features of a memristive material, including short- and long-term memory in a single 2-T device (**Figure 1d**).

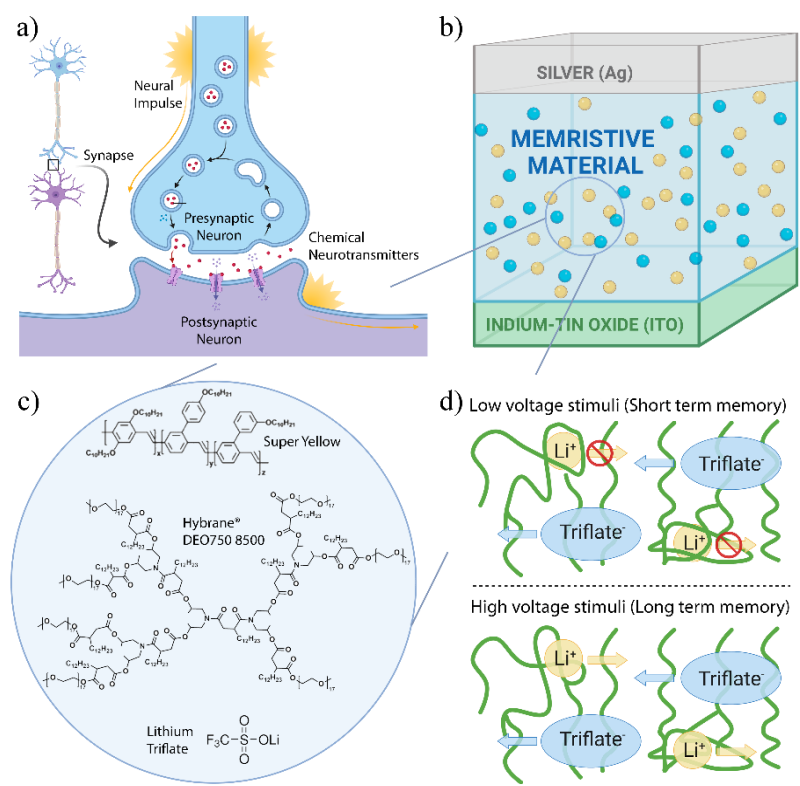

**Figure 1**: (a) Depiction of the electrical transmission in a neural impulse: neurotransmitters are discharged into the synaptic cleft between the presynaptic and postsynaptic neurons, facilitating information transmission. (b) Schematic of a single synaptic device, where, in absence of voltage, the memristive material shows homogeneous presence of ions. (c) Composition of the memristive layer: Super Yellow (semiconductive polymer), Hybrane® DEO750 8500 (ion-transport polymer), and lithium triflate ($LiCF_3SO_3$). (d) Proposed mechanism of the origin of the short/long term memory (STM/LTM) considering the distinct mobilities of ions (blue, yellow) embedded in the polymeric matrix (green).

In a first step, current-voltage (*I-V*) curves were measured to determine its artificial synaptic nature. The current dependence on triangular voltage sweeps applied to the device produced the typical hysteretic behaviour, shown in **Figure 2a**. We found that successive applications of single-polarity sweeps offer an increasing current, meaning that the conductance of the device evolves with the application of voltage in time. This process is repeatable and reversible, achieving different conductance levels, which translates in current increments at certain reading voltages being distinguishable between successive cycles. For example, **Figure 2a** shows that the current increases by 140% from the first to second cycle, and at the 10$^{th}$ cycle, it still increases by 6.3% from the 9$^{th}$ cycle. In **Figure 2b**, a more detailed continuous evolution on the conductive properties is displayed. In this case, the pulse sequence consisted of 50 pulses of 1V, followed by 50 pulses of -2V. It is then validated that the conductance can be continuously increased by at least 200%, offering a wide range of accessible resistance states. This gradual and continuous tuning of the resistance is an advantage with respect to the more common resistance switching devices, where only a few conductance states are achievable in a usually irreversible process.

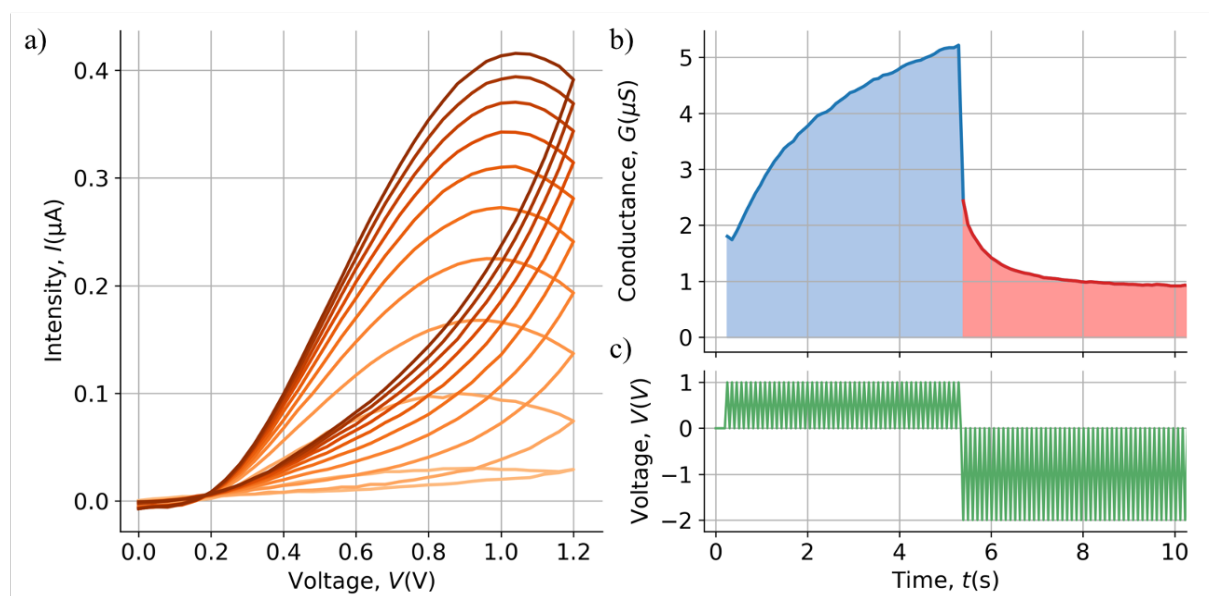

**Figure 2:** (a) Hysteresis loops obtained by successively applying triangular voltage sweeps at a rate of 0.25V/s, waiting 10s between each cycle. (b) Learning/forgetting rate in a single memristive device, achieved by applying the pulses shown in *c*. (c) A fine tuning of the conductance exhibited in *b* can be realized by selecting the pulsed voltage sequence.

Once we demonstrated a fine tuning of the devices' conductance, a second step consisted in unveiling the synaptic memory times of the ionic material. Here, our objective was to determine the changes in conductance obtained by a single pulse and its time evolution by testing the excitatory postsynaptic current (EPSC) (**Figure 3a**). The measurements were applied as follows: first, the device was initiated on its ground state by applying $V_0 = 1V$ during 1s (state $S_0$). Then a short presynaptic positive voltage pulse of 2V was applied during 0.05s to start the EPSC. Subsequently, the system was left at 1V so it spontaneously relaxes to an excited steady state ($S_1$) for 1s. This sequence was repeated to achieve a second excited state ($S_2$). After that, two equivalent postsynaptic negative pulses of -2.5V were applied during 0.05s to obtain the corresponding decreased EPSC, showing states $S_3$ and $S_4$. With this simple experiment, we determined four different, easy to achieve conductive states around ground level. The EPSC ratios, measured as the conductance of an excited state ($S_n$) divided by that of the ground state ($S_0$), are 7.333, 15.553, -3.716, and -16.967 for $S_1$, $S_2$, $S_3$, and $S_4$, respectively. These values are large enough to assure differentiation in order to identify the conductance levels.

Furthermore, two extra experiments were performed to quantitatively determine the tunability of the conductance by applying a particular train of pulses. In the first test, our objective was to determine the plasticity of the material by applying a controlled number of voltage pulses. In the second test, we aimed to determine the relaxation rate of the memory states as function of certain reading time. In **Figure 3b**, a schematic representation of the voltage pulses for both tests is displayed. Here, a starting reading pulse ($V_{read} = 0.5V$) was applied to measure the initial conductance state ($G_0$). This voltage was selected to be low enough to avoid ion migration, and as an indicator of the starting state of the device. Then, a sequence of pulses ($N_{pulses}$) was applied, maintaining $V_{write} = 1V$ and $t_{write} = 0.1s$ (time between each writing pulse). After waiting a variable amount of time ($t_{wait}$) with the circuit set to high impedance conditions, a second reading pulse was applied to determine the new conductance state ($G_f$). In these measurements, a reset time of 300s between experiments was selected, with the circuit now set to short-circuit conditions. This was done in order to guarantee a total recovery of the initial conductance state. In all cases, the variation of the device's conductance was evaluated by the ratio between the two reading pulses, $\Delta G = G_f / G_0$. Each studied variable ($t_{wait}$, $N_{pulses}$) was analyzed independently for either test.

For statistical reasons, several independent devices were measured. The dispersion of measurements was found to be low, especially at a low range of $N_{pulses}$, and for a long $t_{wait}$, demonstrating great reproducibility in a large range of parameters.

From these experiments, we conclude that the conductance of the device is susceptible to be finely controllable, gaining a wide range of achievable states with a very low energy cost (~50 nJ/event). This is illustrated in **Figure 3c**, where $\Delta G$ is linearly represented versus $N_{pulses}$ in logarithmic scale for a waiting time of $t_{wait}$ = 150ms. In **Figure 3d**, the evolution of the conductance ($\Delta G$) versus $t_{wait}$ is represented. In this test, two different excitatory regimes were explored: in the first one, devoted to achieve STM stimulation, $V_{write}$ = 1V with both $N_{pulses}$ = 10 and $N_{pulses}$ = 50 were set (blue, red). For the case of longer memory times, $V_{write}$ = 3V with $N_{pulses}$ = 10 was found to be enough to achieve LTM. Regarding STM, an initial increase of conductance is achieved at very short times ($t_{wait}$ < 500ms). In LTM, the same phenomenon persists until longer times of $t_{wait}$ < 2s. This feature is related with the ionic movement inertia present after the applied train of pulses. However, for longer $t_{wait}$ values we observed the expected exponential relaxation of the conduction. These decays were fitted to a standard exponential equation (see **Supporting Information S3**), obtaining the characteristic times $\tau_S$ = 2.5 ~ 3s for STM and $\tau_L$ = 4.7s for LTM. Quantitatively, the retention time was extracted by analyzing at which point the increase of conductance is negligible ($G_f / G_0$ < 1.05). The obtained value was limited to 10~15s for STM and more than 45s for LTM, thus defining the open work window for sequential stimulation as well as a fine manipulation at very short times.

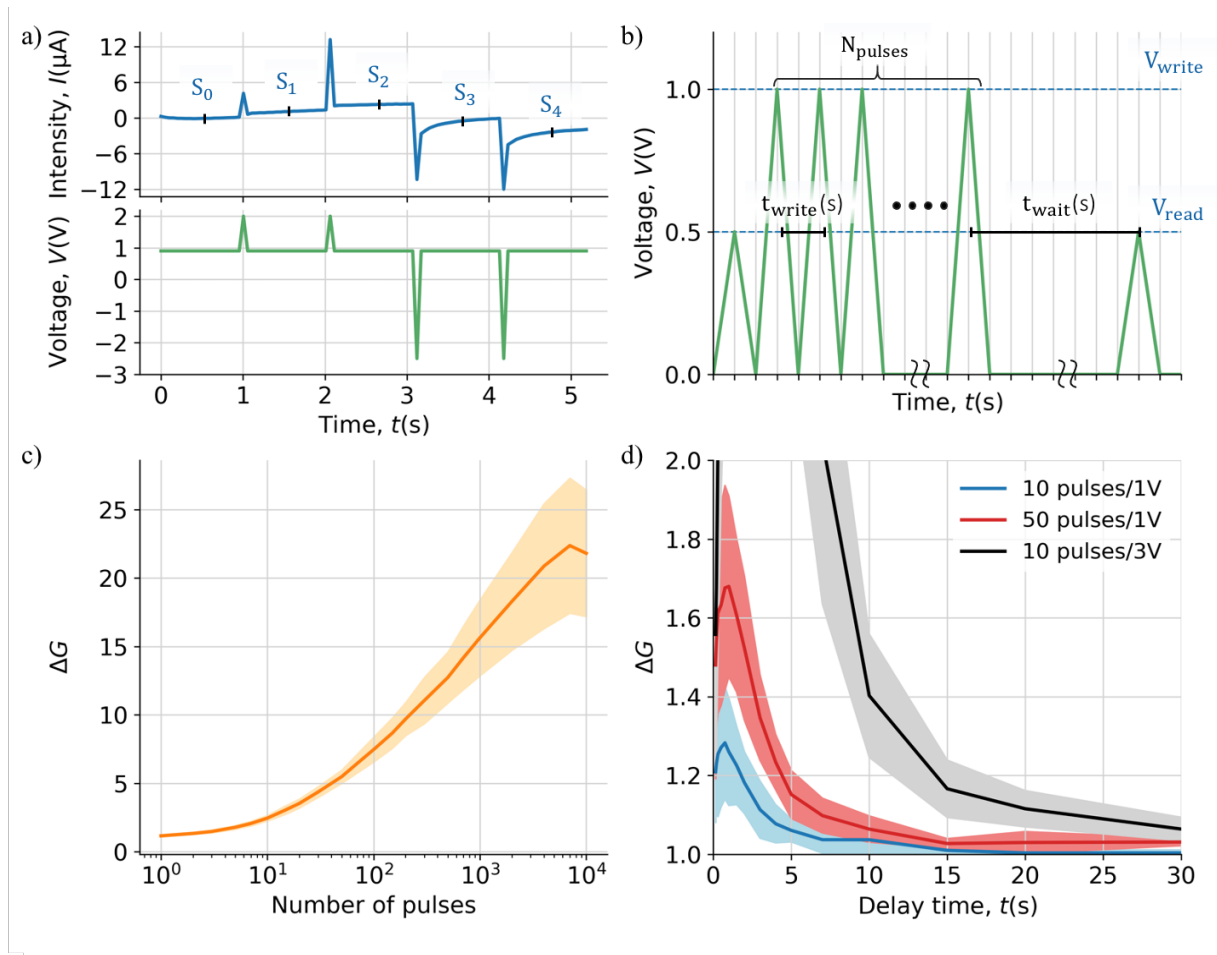

Figure 3: (a) Excitatory postsynaptic current measurements, showing that either positive or negative voltage stimuli produce noticeable variations in the current (and thus, conductance) of the junction. (b) Schematic representation of voltage pulsed trains employed for the experiments shown in *c* and *d*. (c) Determination of conductance susceptibility for a variable number of voltage pulses ($\Delta G = G_f/G_0$); the shadowed area displays the dispersion of measurements. (d) Relaxation rates for short time (blue, red) and long time (black) memory states as function of waiting time.

Moreover, an additional synaptic function called spike-timing-dependent plasticity (STDP), also known as Hebbian learning rule,[43] was tested in these devices. Here, the objective was to translate pulse timing differences into voltage amplitude differences. Experimentally, we confirmed a change in sign and/or magnitude of the conductance with respect to the relative timing between the pre- and post-electronic spikes. This means that the voltage function applied at each point of the experiment was calculated as the voltage function summation between pre- and postsynaptic pulses, which are separated by a delay of *Δt* (**Figure 4a**). With such a disposition of pre- and postsynaptic spikes, the obtained results (**Figure 4b**) correspond to an asymmetric anti-Hebbian learning, which was fitted to the corresponding equation for both positive and negative *Δt* (**Equation 1**):

$$\Delta G = A \exp-\left(\frac{\Delta t}{\tau}\right) + G_0 \tag{1}$$

In both cases, the attained value for $\tau$ is in the range of 85-90ms. This result is in good agreement with the values previously obtained for biological synapses based on real neurons (~100ms).[44]

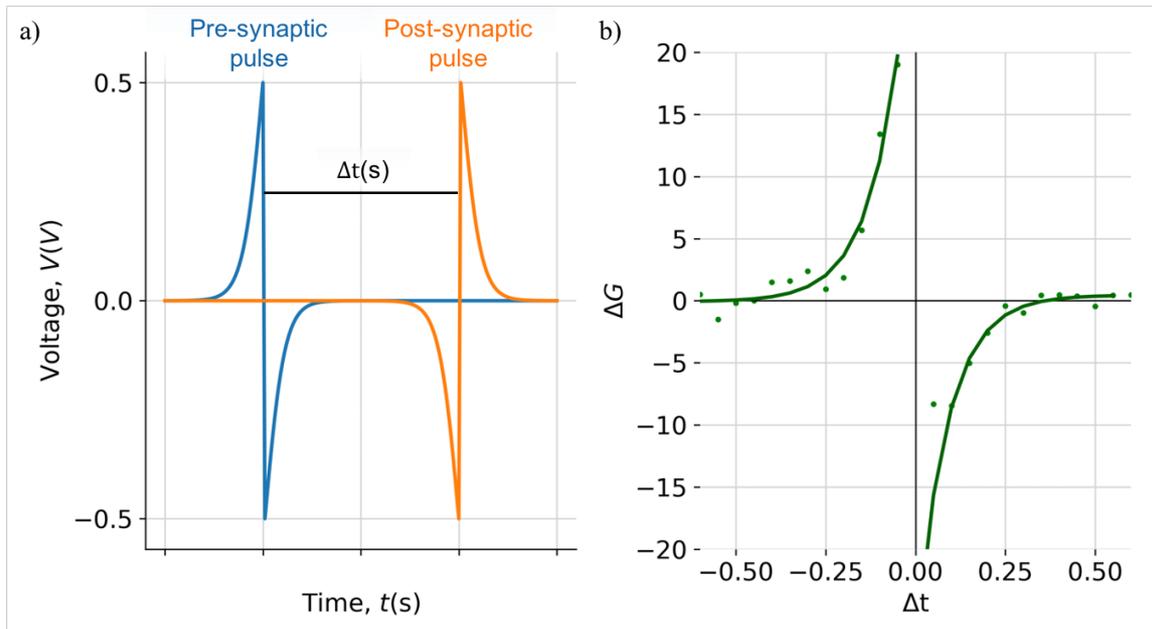

Figure 4: (a) Pre- and postsynaptic spikes separated by a delay of *Δt*, which served as the independent variable in this STDP experiment. (b) Experimentally measured STDP function (fitted to a Hebbian learning model), demonstrating the possibility of translating time-dependent pulses to conductance change.

We estimated that our non-encapsulated memristive devices function properly for two weeks (~300 working hours) after fabrication, even while resting in air at room temperature (see

**Supporting Information S4**), showing an excellent stability for this category of devices[45], but with room for improvement.

3. Theoretical origin of memristance based on ion transport mechanism

In this section, we attempt to describe the observed memristive behaviour by considering a qualitative ionic migration mechanism.[46,47] Initially, in the absence of an applied voltage or current, the distribution of ions in the polymeric matrix is homogeneous and constant along the device. When a certain voltage stimulus is applied, the resulting electric field slightly alters the position of the different anions and cations, modifying the voltage profile of the device and producing a charged double layer in the vicinity of the electrodes. Thus, the nominal resistance of the device can be defined as a function of time, magnitude, and sign of the voltage applied. In this regard, there are two mainly accepted theories to deal with the ionic movement and its effects on the electron injection.[46-51] In the Electrodynamic model,[46,47] an injection-limited regime is considered, and the majority of potential is concentrated in the vicinity of the electrodes. The presence of this ionic layer is known to effectively bend the valence and conduction bands of the polymer, diminishing the Schottky barrier that needs to be surpassed in order to inject electrons and holes. On the other hand, in the Electrochemical Doping model,[48] a reduced voltage drop occurs towards the leads due to the accumulation of ions as in the previous case. However, since the injection is considered to be ohmic in this model, the electronic doping of the polymer is permitted and, consequently, a displacement of cations and anions must occur in order to compensate the doping charges (**Figure 5**). After a strong debate, both models were shown in 2010 to represent different operations depending on the injection regime.[52]

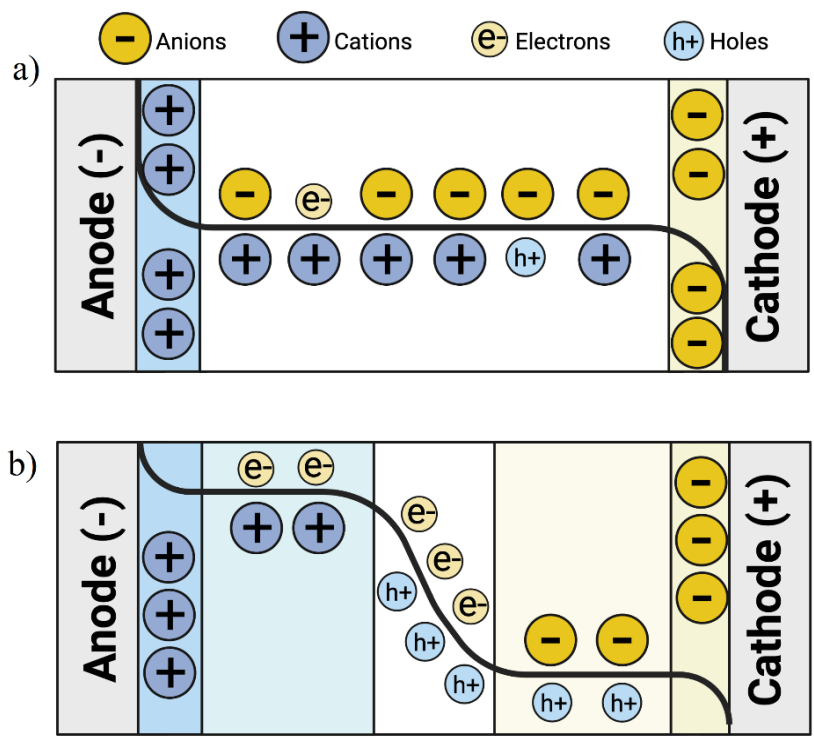

**Figure 5:** Schematic 1D representation of ion migration mechanisms. (a) Electrodynamic model (injection-limited regime; no effective polymer doping exists). (b) Electrochemical Doping model (ohmic injection regime; the polymer conductivity is modified by electronic doping).

The different mobilities in the polymeric environment between fast electrons and holes versus slow anions and cations lead to the fact that, for fast enough measuring times, the charge background and voltage profile can be considered stationary. Thus, the electron/hole injection is adjusted for each static ionic configuration. This could pave the way for modelling out-of-equilibrium states with achievable steady-state approximations.

In our case, the presence of the solid electrolyte polymer mediates both the effective mobility of the ions, and the minimum energy needed to displace them from their equilibrium positions. In this regard, we propose the ionic migration to be the key process determining the remanence times, whose magnitude depend on the interaction between the ion coordinating polymer (Hybrane® DEO750 8500) and the salt ions. Basically, due to acid-base Lewis forces, the Li$^+$ cation (hard acid) strongly interacts with the O atoms (hard basis) present in the Hybrane® structure. Therefore, only high enough voltage stimuli (i.e., 3V) are able to promote the movement of Li$^+$ ions. However, this is not the case for the triflate anions. As a soft basis with less charge/surface, the salt anion is known to interact poorly with oxygen atoms from Hybrane®, thus being easily displaced by a weaker bias voltage.

4. Conclusion

We have reported the material engineering and fabrication of a novel type of organic memristive device based on a conductive polymer-solid electrolyte composite. The electrical characterization in terms of postsynaptic current (EPSC) and neural plasticity (STDP) offered values in close agreement with their biological counterparts. Our contribution states that the Coulomb interactions between the solid electrolyte and solvated ions point towards two ion conduction regimes, which probably are at the origin of the coexistence of STM (10~15s) with LTM (>45s) in a single 2-T device. Using such a simplified structure is a clear advantage in processability and energy consumption compared to previous 3-T attempts in organic devices. Finally, the continuous variation in the conductance of the material has been explained by adapting the theoretical models previously developed for electrochemical cells. Within these mechanisms, memristivity is naturally arising from the continuous displacement of ions, and eventually, from certain polymer doping, depending on the injection regime.

The above results show that our ion-polymer material can be successfully used for the development of artificial synaptic devices in which the conductance of the junctions can be finely manipulated, obtaining a quasi-continuous analogic behaviour with a low power consumption (~50nJ/event) and complete reversibility. This is a key improvement from the previous inorganic contributions where low reproducibility only allows to obtain a digital switchable device with a low number of achievable conductance levels. Yet, while the robustness of these organic devices is higher than those reached in other organic proposals, it is still lower than that of the inorganic counterparts.

To improve the robustness of our devices, further work will be developed based on chemical tailoring. This chemical design is expected to offer, in addition, larger memory times and faster, lower energy-demanding synaptic operations.

5. Experimental Section/Methods

Materials: the semiconducting polymer (Super Yellow, SY), $LiCF_3SO_3$ and cyclohexanone were purchased from Merck, and Hybrane® DEO750 8500 from Polymer Factory. All reagents were used without any further treatment.

Solution Preparation: SY, Hybrane® and $LiCF_3SO_3$ were dissolved in cyclohexanone at a concentration of 15, 10, and 5 mg/mL, respectively, after which the SY solution was left under continuous stirring at 50ºC overnight. They were later mixed following the mass ratio of 1:0.30:0.09 with respect to SY, ending with a final concentration of 8.72, 2.62, and 0.78 mg/mL. This process was performed in a $N_2$-filled atmosphere.

Device Preparation: ITO-patterned substrates were washed ultrasonically in water-soap (Mucasol), water, and 2-propanol baths before putting them inside a glovebox. Then, they were dried with a $N_2$ gun, and the 60s, 2000rpm spin-coating process was performed with the prepared solution. Once the active material had been deposited, the substrates were annealed at 75ºC for 3 hours. Subsequently, 100nm of Ag was evaporated on top of the active layer using a shadow mask. The whole process (excluding the initial substrate washing) was performed in a $N_2$-filled atmosphere.

Profilometry Characterization and AFM Imaging: the thickness of the deposited active layer was determined with the Ambios XP-1 profilometer. The AFM imaging was performed with the Bruker Dimension Icon AFM set in tapping mode. The tip used was of silicon with a resonant frequency of 300 kHz and a force constant of 40 N/m. The scan rate was of 0.6 Hz.

Both studies were performed in air under room temperature. Devices employed for these studies were not electrically characterized.

Electrical Characterization: Devices were electrically characterized using a Keithley 2450 SourceMeter and self-made programming code written in the Keithley Test Script Builder. The devices remained in a $N_2$-filled atmosphere during the whole procedure, except for the case of some degradation studies.

**Supporting Information**
Supporting Information is available from the Wiley Online Library or from the author.

Acknowledgements

The research reported here was supported by the Spanish MINECO (Excellence Unit María de Maeztu CEX2019-000919-M), the European Union (ERC-AdG MOL2D 788222 and ERC AdG HELD 834431), the Generalitat Valenciana (Prometeo Program of Excellence) and the Universitat de València (PRECOMP14-202646). S.C.-S. thanks the Spanish MINECO for a 'Juan de la Cierva-Incorporación' grant, S.G.-S acknowledges the Ministry of Education of Spain (grant PRE2018-083350), C.P.-S. thanks the Spanish MINECO for a Predoctoral Grant (PRE2020-093666). We also thank Denissé de Dios Torralba, Angel López and Alejandra Soriano for their experimental support. Finally, we thank Michele Sessolo for his useful advice about LECs.

Received: ((will be filled in by the editorial staff))
Revised: ((will be filled in by the editorial staff))
Published online: ((will be filled in by the editorial staff))

The table of contents entry should be 50–60 words long and should be written in the present tense. The text should be different from the abstract text.

*Carlos David Prado-Socorro, Silvia Giménez-Santamarina, Luis Escalera-Moreno, Lorenzo Mardegan, Henk J. Bolink, Eugenio Coronado and Salvador Cardona-Serra\**

**Engineering short/long term memory in a molecule-based memristive device.**

ToC figure ((Please choose one size: 55 mm broad × 50 mm high **or** 110 mm broad × 20 mm high. Please do not use any other dimensions))

In this work, we report the memristive properties of a two-terminal (2-T) organic device, based on ionic migration mediated by an ion-transport polymer. Our material can be proposed for the development of artificial synaptic devices in which the conductance of the junctions can be finely manipulated obtaining a quasi-continuous analogic behaviour with a low power consumption (~50nJ/event) and complete reversibility.

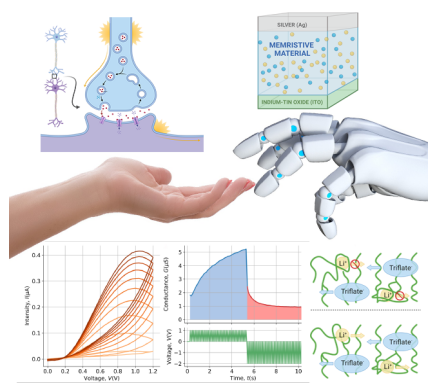

Supporting Information

**Polymer-based composites for engineering organic memristive devices**

*Carlos David Prado-Socorro, Silvia Giménez-Santamarina, Lorenzo Mardegan, Luis Escalera-Moreno, Henk J. Bolink, Salvador Cardona-Serra\* and Eugenio Coronado*

# S1.- Profilometry Characterization

Each sample consists of sixteen different devices that can be studied independently. That is, the finished sample possesses sixteen spots, or cross-sectional areas, where the three layers (bottom electrode, active material, top electrode) coexist. These areas are present in eight rows that can be numbered (Figure S1.1).

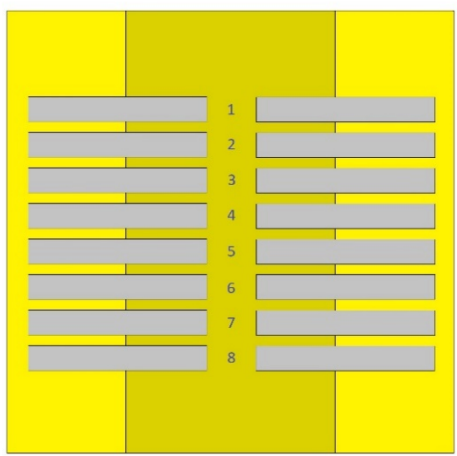

**Figure S1.1:** Schematic of a finished sample with eight rows of two devices each. The spin-coated active material (yellow) is sandwiched between the bottom electrode (ITO layer, darker region) and the top electrode (Ag layer, grey) in sixteen areas (devices).

In order to extract sufficient statistical data, profilometry was performed once for each row of devices, totaling eight measurements per sample. This was done between the Ag segments, so that the scratched area only took the active material layer away. This means that the measurements performed only correspond to the thickness of the active material layer.

The results for a batch of twelve samples, which were spin-coated with the standard setting of 2000rpm (1 minute), are shown in Table S1.1. The equipment utilized was the Ambios XP-1 profilometer.

**Table S1.1:** Recollection of profilometry measurements performed (in *nm*).

|  |  | Samples | | | | | | | | | | | |
|---|---|---|---|---|---|---|---|---|---|---|---|---|---|
|  |  | 1 | 2 | 3 | 4 | 5 | 6 | 7 | 8 | 9 | 10 | 11 | 12 |
| Rows | 1 | 212 | 216 | 205 | 209 | 211 | 246 | 198 | 183 | 197 | 194 | 208 | 214 |
|  | 2 | 205 | 218 | 202 | 207 | 215 | 225 | 193 | 193 | 203 | 204 | 205 | 204 |
|  | 3 | 223 | 216 | 221 | 208 | 224 | 220 | 203 | 203 | 195 | 205 | 216 | 203 |
|  | 4 | 231 | 218 | 213 | 208 | 216 | 220 | 196 | 212 | 205 | 213 | 207 | 199 |
|  | 5 | 220 | 217 | 210 | 213 | 209 | 212 | 206 | 216 | 193 | 206 | 214 | 221 |
|  | 6 | 238 | 216 | 208 | 211 | 218 | 221 | 198 | 206 | 199 | 197 | 212 | 204 |
|  | 7 | 204 | 219 | 208 | 214 | 207 | 212 | 198 | 203 | 203 | 207 | 200 | 210 |
|  | 8 | 225 | 217 | 207 | 207 | 206 | 201 | 205 | 189 | 197 | 196 | 201 | 221 |
| Average | | **209 ± 10.2 *nm*** | | | | | | | | | | | |

## S2.- AFM Imaging

The topography of the spin-coated substrate was imaged using a Bruker Dimension Icon Atomic Force Microscope set in tapping mode The tip used was of silicon with a resonant frequency of 300 kHz and a force constant of 40 N/m, while the scan rate was of 0.6 Hz. Six 5x5µm images were taken for a sample spin-coated at 1750rpm for 1 minute (Figures S2.1-S2.6). These images were extracted employing the *Gwyddion v2.58* software.

AFM images of the sample showed the absence of huge aggregates. However, small features are observable, corresponding to bubble-like formations or holes that may form during the annealing process. Additional experiments are suggested in order to study the effects that different annealing times and temperatures, spin-coating speeds, or ratios of components present in the active layer, can exert on the topography characteristics of the finished samples.

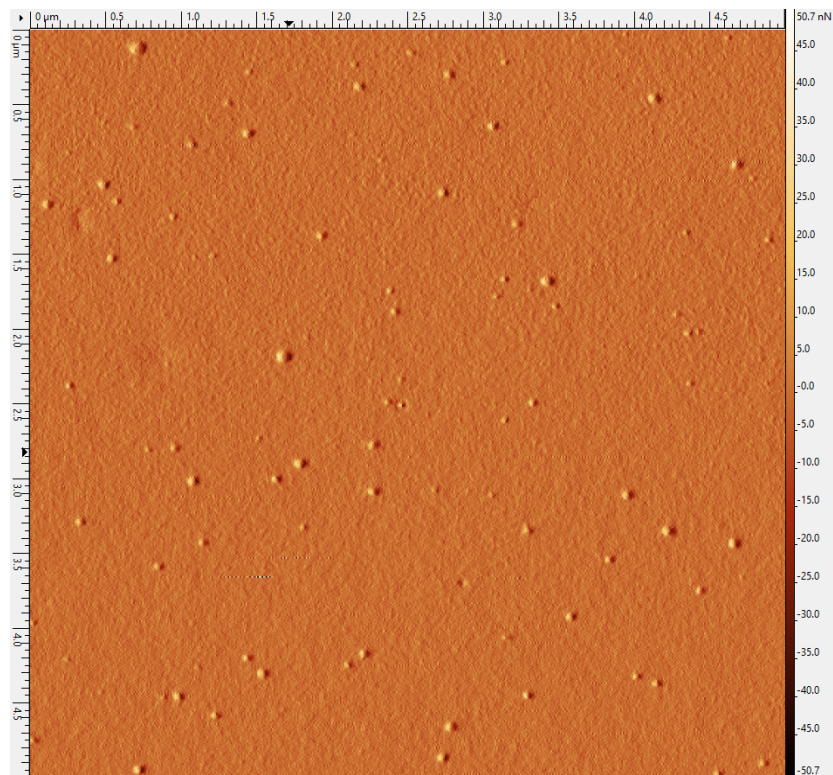

**Figure S2.1:** AFM, 1750rpm (1).

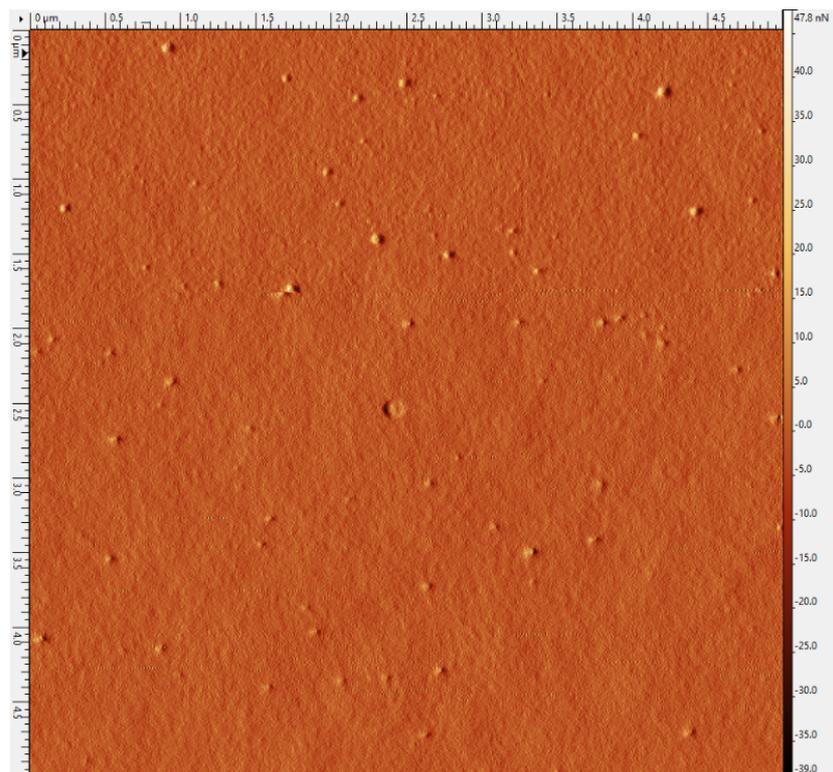

**Figure S2.1:** AFM, 1750rpm (2).

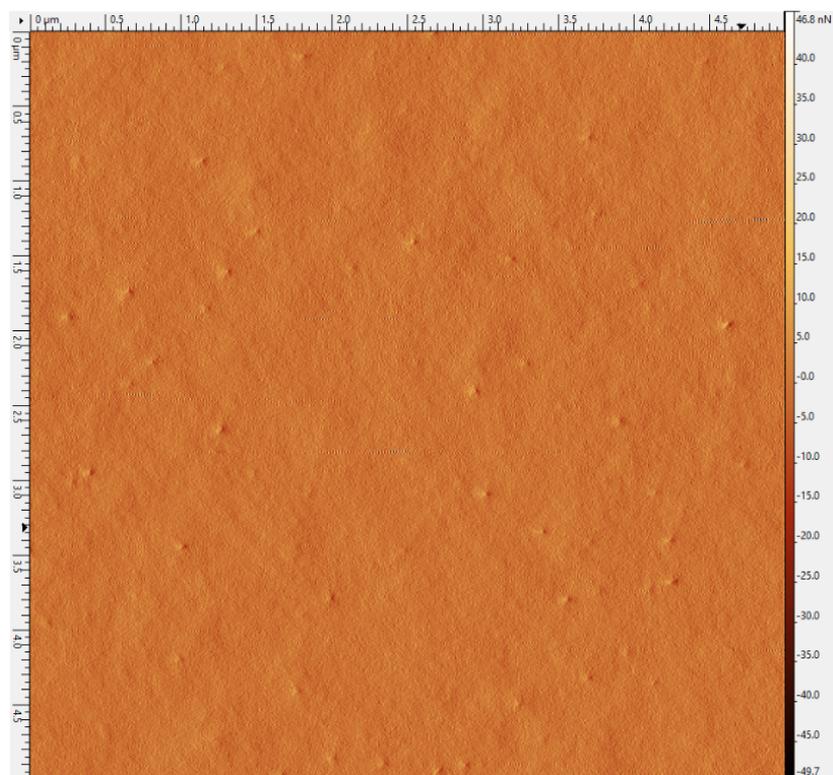

**Figure S2.2:** AFM, 1750rpm (3).

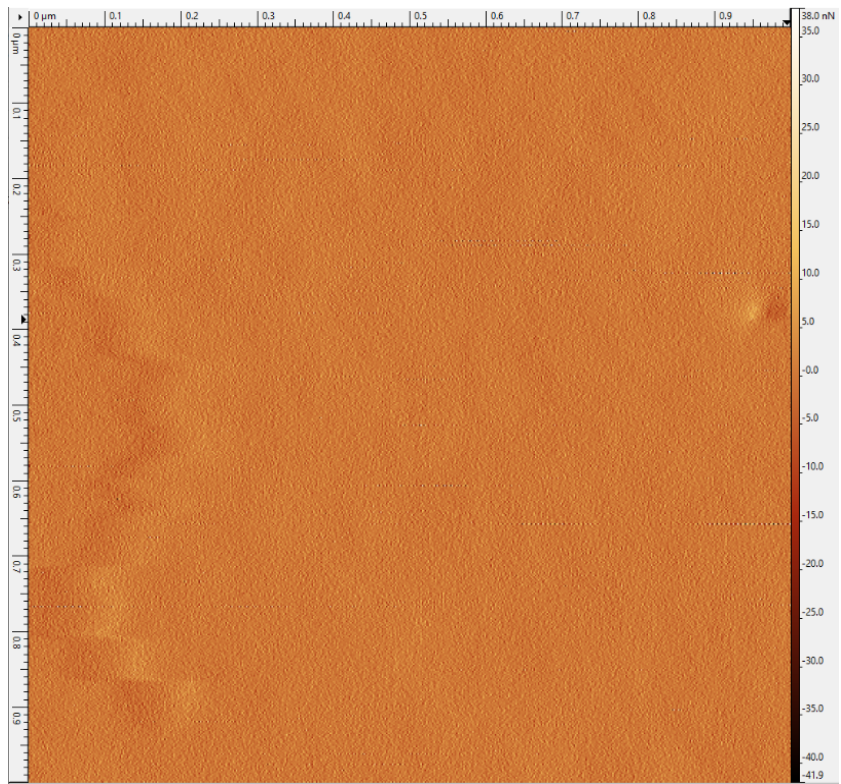

**Figure S2.3:** AFM, 1750rpm (4).

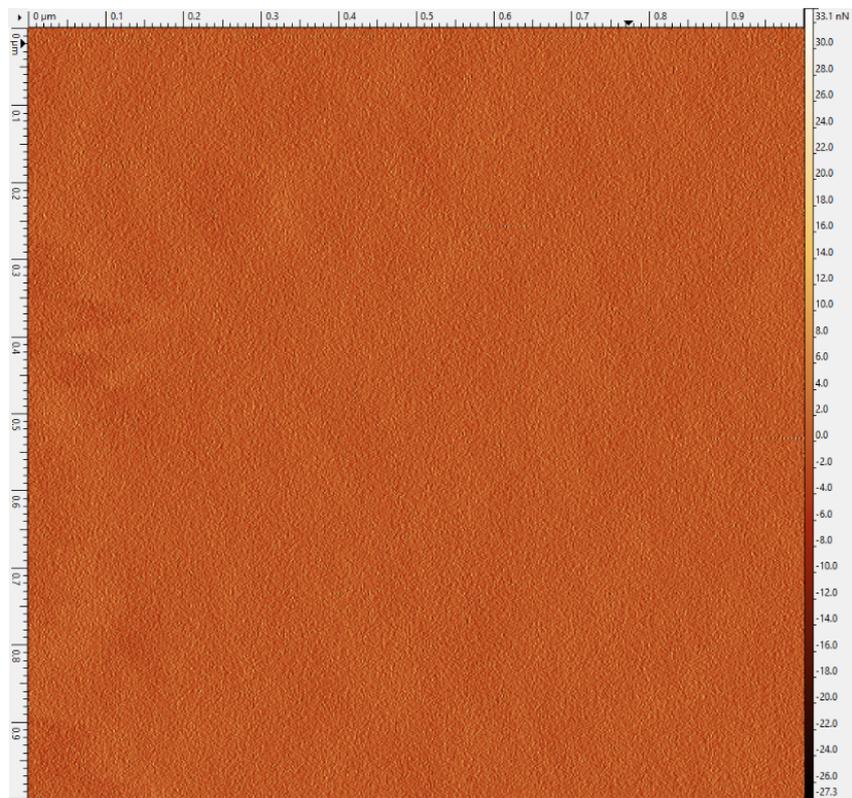

**Figure S2.4:** AFM, 1750rpm (5).

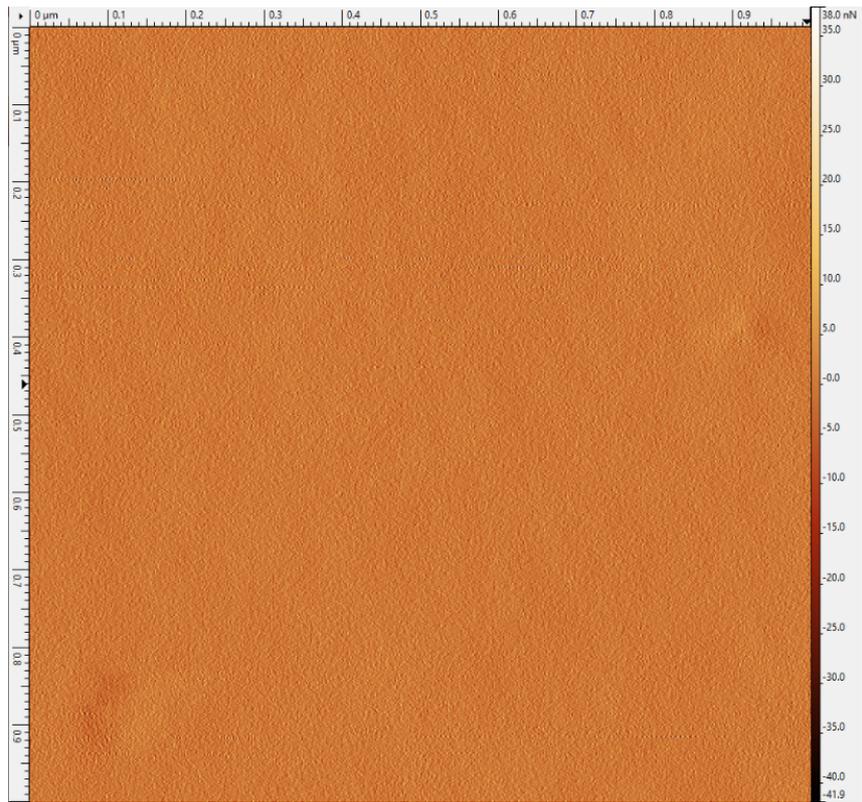
**Figure S2.5:** AFM, 1750rpm (6).

## S3.- Exponential Fitting of the STM and LTM decays

In **Figure 3d** (main document), the evolution of the conductance ($\Delta G$) versus $t_{wait}$ is represented. In the experiment, we observed that, for $t_{wait}$ values longer than 500ms (2s for 3V excitation), the expected exponential relaxation of the conduction was present. These decays were fitted to a standard exponential equation (S3.1) using a minimum squares method as implemented in the program *IGOR PRO 8*. (Figure S3.1):

$$G_f/G_0 = y_0 + Ae^{t/\tau} \qquad (S3.1)$$

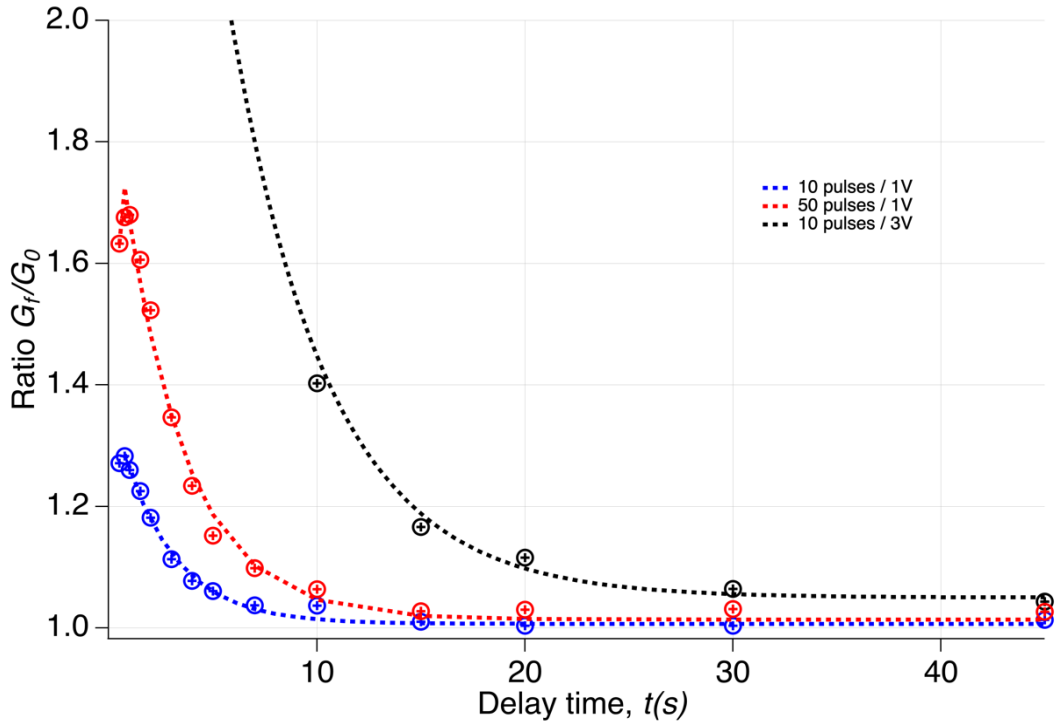

**Figure S3.1:** Fitting of the relaxation rates for short time (red, blue) and long time (black) memory states as functions of waiting time (dashed line: fitting; dots: experimental data).

The parameters resulting from the empirical fitting are summarized in Table S3.1. In all the cases, the $y_0$ parameter value must be ~1 as it represents the minimum value for the $G_f/G_0$ ratio. Besides, the parameter $A$ can be ascribed to the inertia felt by the ions due to the potential bias gradient. Finally, regarding the characteristic times, which shapes the curves and gives an insight of the relaxation procedure, we obtained $\tau_S$ = 2.5 ~ 3s for STM and $\tau_L$ = 4.7s for LTM.

**Table S3.1:** Summary of fitting parameters obtained from the STM / LTM experimental tests.

| Experimental Test | $y_0$ | $A$ | $\tau$ (s) |
|---|---|---|---|
| 10 pulses / 1V (*) | 1.01 | 0.374 | 2.57 |
| 50 pulses / 1V (*) | 1.01 | 0.91 | 3.01 |
| 10 pulses / 3V (*) | 1.05 | 3.30 | 4.73 |

## S4.- Degradation vs. time in air and in a $N_2$-filled glovebox

Figure 2a (main document) shows the most common electrical characterization technique that was performed in this study in order to discern between functioning and broken devices, and to get a rough estimate of their conductivity. This type of measurement was performed with the Keithley 2450 SourceMeter, along with self-made programming code written in the Keithley Test Script Builder.

The hysteresis curves themselves consisted of ten voltage sweeps between 0 and either 1.2 or 3V for 'low' or 'high' voltage studies, respectively. For the case of studying degradation vs. time, a 'high' voltage setting was used, since it represents the most demanding test for the devices, and it involves internal processes happening within the active material (i.e., movement of $Li^+$ cations). The rate of sweeps was kept at 0.62 V/s, with a 10s delay between each cycle.

A non-encapsulated sample containing sixteen independent devices was left outside of the $N_2$-filled glovebox in order to test its robustness against room conditions. Two devices were measured once every other day until Day 14 after fabrication, in the following way: at Day 0 (just after evaporating the Ag electrode on top), a pair of devices was characterized; at Day 1, this pair was also measured, along with a new pair of devices from the same sample, and so on. This was done in order to differentiate between changes caused by excitation, and those caused by time degradation (Figure S4.1). The incidence of light was not controlled for this sample.

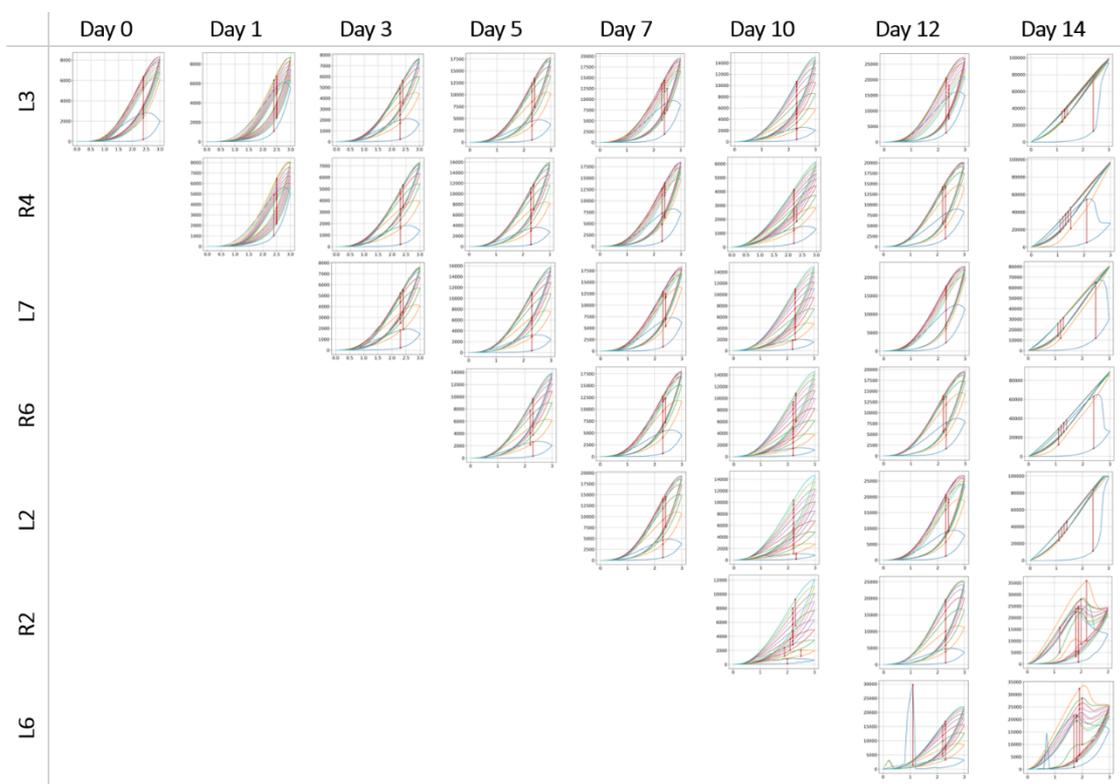

**Figure S4.1:** Degradation vs. time study of multiple devices in a single non-encapsulated sample, which remained in air at room temperature. Columns represent days after fabrication. Hysteresis curves demonstrate correct functioning of the devices until Day 14, when all of them simultaneously show an erratic behaviour. Measured devices not shown in the figure present similar or identical curves.

We found that changes caused by excitation are negligible at this days-long time scale, since devices measured for the first time in the following days display similar or identical curves than those that were measured previously. Furthermore, the devices showed nicely-shaped hysteresis curves that increased in conductance levels every successive cycle until Day 14, after which they stopped functioning properly. This suggests that our devices can work in air and under room temperature for around 300 hours.

Additionally, a pair of non-encapsulated samples remained inside a $N_2$-filled glovebox to test its robustness against time, rather than air. Sample 1 was kept in the absence of light, while Sample 2 remained with the light turned on at all times. Both of them exhibited a perfect behaviour even at the 14$^{th}$ day after fabrication (Figure S4.2).

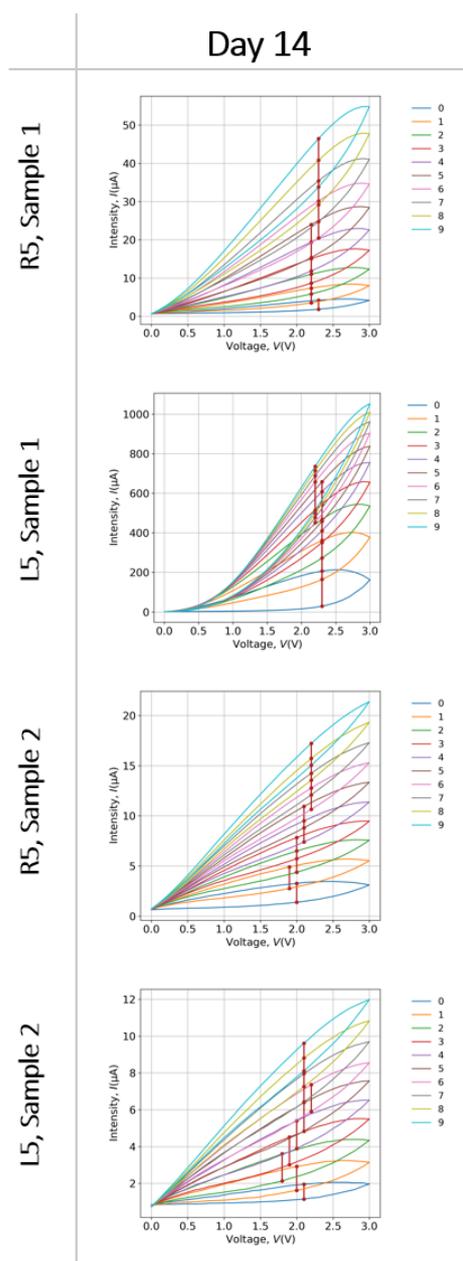

**Figure S4.2:** Hysteresis curves of devices kept inside a $N_2$-filled glovebox for two weeks after fabrication. Sample 1 was kept in the absence of light, while Sample 2 remained with the light turned on.

No discernible, clear differences were found between devices kept within light and those kept in the dark, although further experimentation is suggested in this regard to discover the optimal conditions for storing the devices, as well as their real lifetime. Moreover, given the more aggressive nature of the electrical characterization performed in these experiments, where a changing voltage is applied for 10s per cycle, with just 10s between each one to relax, some devices may deteriorate, showing much higher currents. Thus, the dispersion from the group in this experiment is higher compared to the softer pulsed voltage studies displayed in the main document. Overall, the robustness of our non-encapsulated devices was found to be high, especially compared with standard organic devices. However, further chemical design is being developed to improve this in future iterations.